\pdfoutput=1

\documentclass[11pt]{article}
\usepackage{jheppub}

\usepackage{bm}

\usepackage[usenames,dvipsnames,svgnames,table]{xcolor}

\usepackage{amsmath}
\usepackage{amssymb}
\usepackage{graphicx}
\usepackage{grffile}    
\usepackage{slashed}
\usepackage{soul}
\usepackage{multirow}
\usepackage{pdflscape}
\usepackage[section]{placeins} 
\usepackage{lineno}

\usepackage{mathtools}
\usepackage{siunitx}
\DeclarePairedDelimiterXPP\BigOSI[2]%
  {\mathcal{O}}{(}{)}{}%
  {\SI{#1}{#2}}

\graphicspath{{plots/}}

\title{
\begin{center}
Readout Technologies for Future Detectors
\end{center}
}

\author[b]{M.~Begel,}
\author[c]{J.~Eisch,}
\author[d]{M.~Garcia-Sciveres,}
\author[a]{A.~Paramonov,}
\author[a]{M.~Trovato,}
\author[a]{J.~Zhang}

\affiliation[\ ]{\quad}
\affiliation[a]{Argonne National Laboratory}
\affiliation[b]{Brookhaven National Laboratory} 
\affiliation[c]{Fermi National Accelerator Laboratory}
\affiliation[d]{Lawrence Berkeley National Laboratory}

\abstract{
\newline 
It is essential to develop novel cost-effective readout technologies to maximize the discovery
potential of future HEP experiments.
}

\vfill
\break 
\vfill
\break

\begin{document}

\begin{flushright}
\small{.}
\end{flushright}

\maketitle
\flushbottom

\medskip
\section{Introduction}
\label{s.intro} 

\par The field of particle physics needs to continue developing and adoption of novel trigger and data acquisition (TDAQ) technologies for the future experiments, to be able to cope with extremely large instantaneous data volume in harsh environment, to optimally examine the data collected, and to optimize the cost and usage of off-detector resources. The Large Hadron Collider (LHC) is expected to start the high luminosity (HL-LHC) operation era in 2029, to ultimately reach the peak instantaneous luminosity of $L = 7.5\times 10^{34} cm^{-2}s^{-1}$, corresponding to $\sim$200 inelastic proton-proton collisions per bunch crossing (pileups), and to deliver more than ten times the integrated luminosity of the Runs 1-3 combined. Moreover, the proposed FCC-hh would be capable of up to $L = 3\times 10^{35} cm^{-2}s^{-1}$ with $\sim$1000 pileups. 

\par For these experiments, an important challenge is to extract the data at high bandwidth from the detector subsystems, particularly the tracking detectors, without adding an excessive burden of material due to the large number of data electrical cables, power delivery, and cooling. The electrical data cables are used to readout the tracking detectors because the modern fiber-optical transceivers are not sufficiently radiation tolerant to be placed inside the tracking detector. High instantaneous luminosity and high occupancy of high-granularity detectors for FCC-hh experiments would yield huge data rates, such as $\sim$800 TB/s from the tracking detectors and $\sim$200 TB/s from a calorimeter combining  electromagnetic and hadronic compartments. 
\par Versatile Link+, a custom radiation tolerant fiber-optical transceiver for the HL-LHC experiments, operates at 10.24 Gb/s per fiber~\cite{CERN:VL, Huffman_2014}. Without significant advances in radiation tolerant data transmission technologies, more than one million optical fibers operating at 10 Gb/s will be needed to transmit 10 Pb/s off an FCC-hh detector to event building network. This would have significant implications on the detector cost and design because of the high mass of passive material for readout, power delivery and cooling services. The low link speed also increases the cost of the off-detector readout since the Multi-Gigabit Transceivers in FPGAs are designed for link speeds well exceeding 10 Gb/s and the number of transceivers in an FPGA is limited. 

\par For future neutrino experiments such as DUNE, and the next generation dark matter experiments such as LZ and SuperCDMS, the volume of data to be produced will be at PB scale as well, which will require significant readout infrastructure. Some of the on-detector data links need to operate in cryogenic environment.

\subsection{Readout Link Needs}

\par The trigger and data acquisition systems are designed to maximize the discovery potential of the experiments given all the technical and financial constraints. Some experiments, like ALICE and LHCb, are read out for every collision without buffering data on-detector and a hardware-based fixed latency trigger. These experiments use commodity computing to filter and process data further. This approach allows development of sophisticated data-processing algorithms, easier maintenance, and straight-forward upgrades. 

\par In contrast with ALICE and LHCb, selected sub-detectors of the ATLAS and CMS experiments can not be readout for each bunch crossing because of the limited readout bandwidth. The bandwidth limitations in readout of the inner trackers come from the radiation tolerance of the optical links and the available space for the electrical data cables. The VCSEL-based optical transmitters are not sufficiently radiation tolerant to be inside the tracking detectors so the transmitters received data via long electrical cables that limit the readout bandwidth. The data rate is about 10-40 times higher than the available bandwidth. The bandwidth limitation mandates on-detector buffering and selected readout of the experimental data using fixed-latency trigger systems. The trigger systems are built with custom hardware and digital algorithms (firmware). They require expert engineers to design and maintain them. 

\par As we discussed earlier, the ability to readout an entire detector at the full collision rate can greatly simplify the TDAQ system and expand the discovery potential. Availability of radiation-tolerant fiber-optical links is one way to solve the problem. The other approaches include techniques such as on-detector data compression, reduction, and aggregation. The data reduction approach intelligent algorithms that remove unwanted signal like noise. For example, the upgraded CMS tracker will output stubs, segments of track trajectories. The stubs are formed with hits from adjacent tracker layers. 

\par The readout links for the future collider experiments need higher bandwidth and radiation tolerance than the present generation. They can also benefit from wavelength division multiplexing (WDM) to reduce the number of fibers between the detector and the readout system and to simplify the on-detector aggregation electronic. 
The on-detector electronics (ICs) for data aggregation, serialization, and transmission need to have comparable radiation tolerance and bandwidth as the fiber-optical data links.


\subsection{Industry Trends and R\&D}

\par The commercially-available link technologies can provide very high bandwidths, and extend to distances of a few kilometers. High data rate is being achieved (800 GbE by Ethernet) via WDM or modulation levels (PAM). VCSEL-based links are still the majority with highly optimized multi-mode fibers, but silicon-photonics based single-mode systems are promising for high-rate and long-distance interconnects. Modern commercial laser-based transceivers operate at 50 Gbps per optical link in 400G QSFP modules with power consumption of about 30 pJ/bit~\cite{Finisar_transceivers}. Comparable 400G QSFP Si-pho transceivers from Intel operate at 100 Gbps per fiber and consume about 20\% less power~\cite{Intel_transceivers}. Intel hopes to scale its silicon photonics platform up to 1 Tb/s per fiber at 1pJ of energy consumed per bit, reaching distances of up to 1 km.

\par Co-packaging of optoelectronics with FPGAs is emerging, while most designs still rely on pluggable optoelectronics. Silicon photonics may be a game-changing technology because of its good integration density and integration synergies with microelectronics. Therefore for HEP applications, COTS components may meet the bandwidth needs but likely custom front-end modules compatible with commercial standards will still be needed.

\par Wireless technologies may provide unique opportunities for future HEP experiments, to control and configure detectors with broadcasting, and to remove partially or completely all the cables and connectors in the detectors. In order to be applicable, high performance transceivers for wireless transmitters and receivers for wave bands suitable for high data rate transmission need be developed , and more importantly, be scalable. Such a wireless data transmission system would need operate in the strong magnetic field, the radiation created by the beam, the RF noise/interference created by neighbor cells, the signal cross-talk and multi-path propagation on metallic parts of the detector, and the variations of temperature and humidity.

\par The interconnect between the custom on-detector systems and commodity computing is the core of data acquisition systems. Modern DAQ systems  for HEP use FPGAs to accommodate custom synchronous protocols for the on-detector systems. The synchronous data transmission is needed to time the digitization of the detector signals to the colliding bunches of particles. Most likely we will have to continue using FPGAs for detector readout because the commodity computing platforms are inherently asynchronous. The connectivity between FPGAs and commodity computing is likely to go via  PCIe interface (Gen 4+). There are attempts to bring network connectivity (e.g. TCP/IP) directly to FPGAs but these designs are either incomplete or slow. The complete TCP/IP designs utilize built-in microprocessors which are slow and limit the bandwidth. There are other board-level interconnect technologies such as Infinity Fabric and Ultra Path Interconnect but they are not naively supported by the modern FPGAs.

\par There may be future opportunities to use FPGA-CPU hybrids of PCIe FPGA boards. For example, Intel Xeon Gold 6138P with Arria 10 FPGA instead features a High Speed Serial Interconnect (HSSI) bus with 16 serial 10+ Gb/s lanes and it comes at a fairly competitive price increment of about \$2300 in comparison to a regular CPU. Unfortunately, these serial links are not fully customize and there are no mechanisms to supply user-defined clocks to the FPGA. These issues prevent us from using the hybrid CPU for detector readout. This particular CPU may be beneficial for low-latency (or high throughput) custom computing algorithms. Nevertheless, future hybrid CPUs may become suitable for detector readout as Intel and AMD continue competing.

\par The newer TDAQ systems require more and more effort to design and verify. This trend is driven by the increase in complexity of the on-detector readout systems, FPGAs, interconnect technologies (e.g. PCIe Gen 4), and system specifications. The development effort is dominated by the testing and verification, not firmware or software design. Testing and verification with simulated data and front-end emulators do help but they can not replace testing with actual detector systems where we encounter undocumented features. Therefore, we may benefit from novel design and verification approaches to reduce the total design effort.


\section{Technologies}
\label{s.technologies} 

\subsection{High-speed Rad-hard Links}
\label{s.opticallink} 


\par Bringing fiber-optic interconnects directly to a tracker sensor module or to the sensor readout IC can enable triggerless readout of of collider experiments and the data transmission via long electrical cables would not be needed. The optical fibers offer much higher bandwidth, lower mass and power consumption, and easier routing in comparison to the contemporary electrical interconnects. The electrical to optical conversion can happen on the same PCB or flex cable where the sensor readout ICs are installed (fiber-to-module). The electrical-to-optical conversion can also be integrated into the sensor readout IC (fiber-to-chip).

\par The Silicon Photonic technology promises fiber-to-module and fiber-to-chip readout~\cite{9380443}. It is a mature commercial technology. Si-Photonic transceivers are  developed by Intel, Cisco (Luxtera), Broadcom, and other companies. Contemporary commercial Si-Pho products offer about 100 Gb/s per fiber with power consumption of about 1.5W (that includes power for the CW laser)~\cite{Intel_transceivers}. In HEP experiments the laser can deliver light off-detector to the optical chip through a fiber. Si-photonic offers low power consumption, high-reliability, small footprint, low cost, and customizable radiation tolerance. Si-Pho optical circuit can be fabricated in the same wafer together with the electrical circuits using a conventional CMOS process or separately (and bump-/wire- bonded to the sensor readout chip)~\cite{doi:10.1063/5.0050117}. 

\par Commercial Si-Photonic devices have been evaluated for radiation tolerance if HEP experiments and the optical circuits themselves were found to be highly rad-tolerant~\cite{Kraxner:2019Il,Prousalidi:2781875}. of The optical components can be designed to be highly radiation tolerance. The commercial devices have limited radiation tolerance mostly because of the slow-speed electrical interfaces needed for configuration and control (i.e. 3.3V I2C bus). The high-speed components were found to be much more rad-tolerance than the low-speed interface.  So far the majority of fiber-optical data transmission in HEP experiments are relying on VCSELs and EELs~\cite{Huffman_2014, Zeng_2017}. The Si-photonics technology is relatively new to HEP; it has been explored since 2011 for use in the present and future experiments~\cite{Drake_2014, Kraxner:2019Il}.

\par The data aggregation from multiple sources into a single fiber-optical link can be done logically (e.g. like in lpGBT~\cite{TWEPP2019:lpGBT, CERN:lpGBT}) or optically using Wavelength Division Multiplexing (WDM). In the WDM approach, each serial link  is transmitted using its own wavelength. The WDM approach does not require any digital data aggregation but it requires more Ring Resonator Modulators than the other approaches. The WDM data will also need to be de-multiplexed off-detector so commodity receivers can not be used. 

\par The WDM approach allows to avoid bit rate up-conversion and to collect data from distant modules onto the the same fiber. One example, based on SBIR-supported development by Freedom Photonics, LLC, all complex electronics are outside the radiation environment and only modulators on silicon photonic chips must reside in the detector volume. Furthermore, this method does not require a specific bit rate or synchronization for the signals to be multiplexed, even allowing signals with different bit rates to be combined on the same fiber. The footprint and power in the radiation environment are expected to be small, as there is no data aggregation ASIC inside the detector. There is no hard limit on the number of signals that can multiplexed onto one fiber, permitting evolution to ever higher bandwidth detectors.

\begin{figure}[tbp]
\begin{center}
\includegraphics[width=0.8\textwidth]{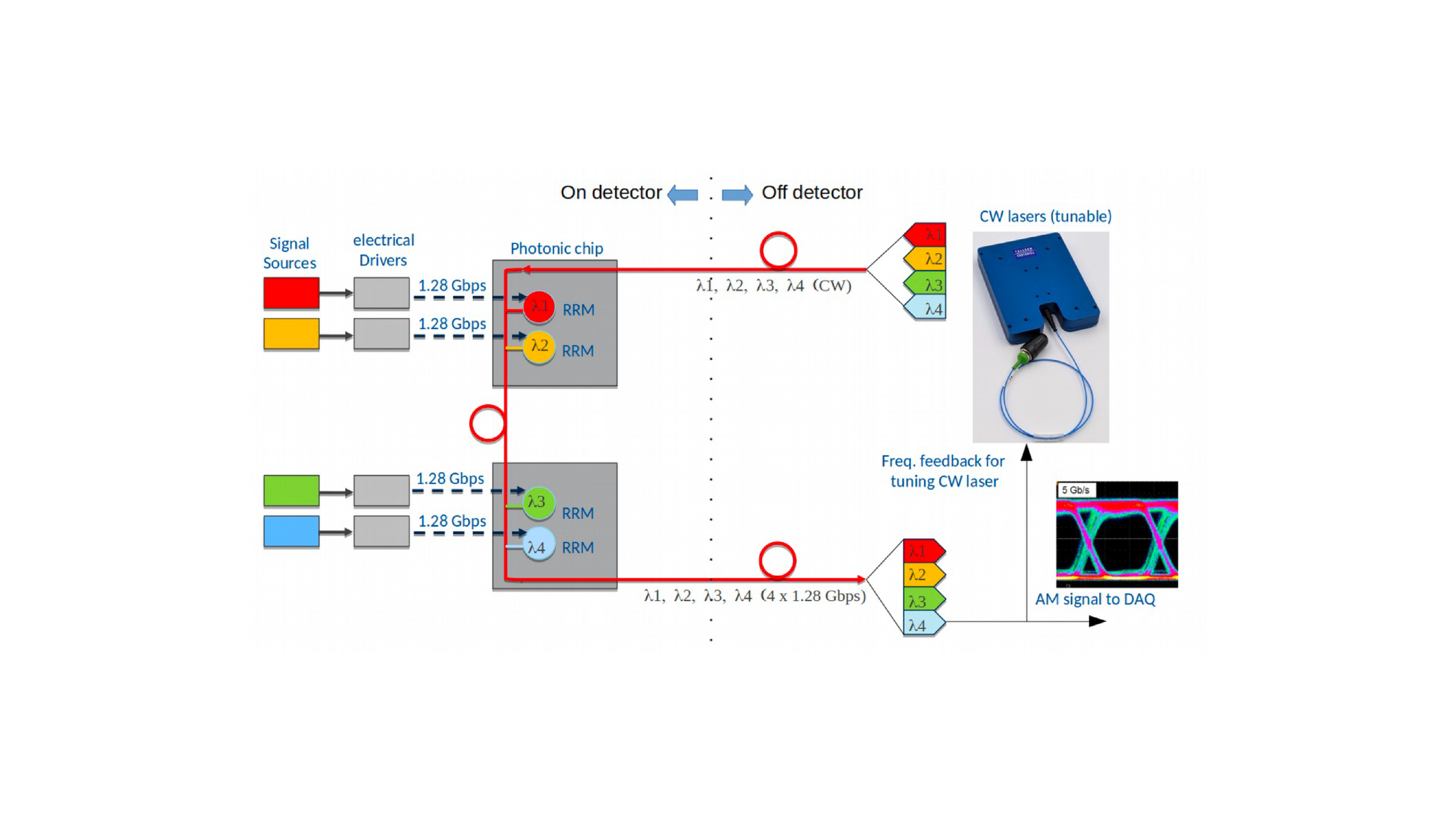}
\caption{WDM readout system proposed architecture. RRM stands for Ring Resonator Modulator\label{fig:WDM_RRM}.}
\end{center}
\end{figure}

\par The proposed concept is shown in Fig~\ref{fig:WDM_RRM}. Multiple laser wavelengths are combined onto a single fiber off-detector. Each wavelength is a CW (Continuous Wave) carrier that is modulated on-detector using a ring resonator. While Current technology is already demonstrating 20~Mbps modulation, the radiation hard data sources (detector front end chips) are expected to remain at the 1~Gbps level, as this simplifies radiation tolerant design, and a typical detector consists of many “low” bandwidth sources rather than few high bandwidth ones. The off-detector tunable lasers can be purchased from Freedom Photonics or somewhere else. 

\par The on-detector amplitude modulation is carried out by Ring Resonator Modulators (RRM) on a photonic chip. These must be radiation hard, and such technology is being developed with SBIR and other support. The electrical drivers to feed the RRM’s are analog circuits that can be separate CMOS chips or possibly integrated into the detector readout chips. As RRMs are also implemented in a (silicon photonic) CMOS process, there is also potential to integrate drivers and RRM’s on the same chip, further reducing mass, complexity, and cost, while increasing performance. Commercial demodulating technology (off-detector) also exists. The Fig.~\ref{fig:WDM_RRM} concept includes distant detector elements multiplexed onto the same fiber. This would allow a macro-assembly to have just one input and one output fibers, with the total bandwidth required dictating how many laser frequencies are needed. Each RRM works at a specific wavelength, but that wavelength can drift, for example due to temperature. The incoming wavelength must track these shifts, so a feedback mechanism is needed. The architecture allows for this by monitoring the return signals on the output fiber. The feedback can be implemented either by adjusting the laser wavelength or by acting on the RRM (for example with a heater). The method used will depend on practical considerations of a given implementation. 


\par VCSEL-based optical interconnects are mainstream in the HEP experiments and should be considered further. However, the HEP community can not re-design VCSELs for radiation hardness while the Si-photonic modulators can be fully customized. The modern commercial VCSELs reach about 50~Gb/s per fiber and their power consumption per bit is only 20\% higher than of the Si-photonic alternatives.

\par The HEP community may benefit from strengthening the high-speed interconnect R\&D program and by taking advantage of the commercial developments. CERN and the European HEP community are already actively exploring the Si-photonic and other technologies with the US HEP effort being behind.

\subsection{Wireless Data Communication}


\par In order to unleash the full potential of the experiments, the available bandwidth must be boosted by increasing the number of links or their transfer rate, or data must be pre-processed on-detector. An example of efficiently pre-processing the data is to enable inter-layer communication in HEP trackers in order to correlate spatial information across different layers and extract particle momenta. The data volume from the detector frontends can be significantly reduced by selecting high-momentum charged particles. Wireless microwave and free space optics technologies can serve for this purpose. Both approaches are describe below.

\subsubsection{Wireless Microwave Technology}
Wireless microwave technology, which is constantly in our daily lives (e.g: transferring of uncompressed data for HDTV), can deliver rates in the range of several gigabits per second, thanks to the constant progress being made. Because of that, microwave technology is a natural candidate to tackle both the need for more links and for the efficiently pre-processing of the data, by allowing to easily share information across the detector. Wireless links consist of antennas and transceivers. A recent design exploiting the 60 GHz band~\cite{RBrenner} was proven to be suitable for HEP detectors. Such a design is 140 times smaller than commercial 5 GHz devices. The transceiver is designed in 90 $nm$ CMOS and the size of the chip is of 2.75~mm $\times$ 2.5~mm. The size of a surface integrated waveguide (SIW)-based four-by-four slot array antenna is approximately 1 cm$^2$. The choice of the antenna is vital to reduce the power consumption and, depending on the antenna directionality, the cross talk. The power consumption of the transceiver are measured to be respectively 170~$mW$ in transmit mode and 135 $mW$ in receiver mode. The measured bit error rate over 1 m distance at 4~Gb/s is $10^{-11}$. Performed tests show the the transmission of data with wireless through one or several silicon layers is not possible, thus making the cross talk problem reduced.

Given the rapid progress in technology, the smaller size of its components and their reduced sensitivity to mechanical damages, the limited cross talk, and the possibility to introduce on-detector intelligence, wireless technology offer an alternative high-rate data transfer to what is already available in the HEP world. The reduced data transmission range because of the high frequency is not a limiting factor, given the short distances involved in the tracker environment. Efforts on wireless techniques are described in the WADAPT (Wireless Allowing Data and Power Transmission) consortium's paper~\cite{WADAPT}.

\subsubsection{Free Space Optics Technology}
Microwave technology has established a strong momentum from their heavy investment and has managed to fulfill the requirements (e.g: 5G infrastructure) up to now. However, microwave-based solutions are reaching their upper physical limit in supporting higher bandwidth for future applications. Higher frequency bands, such mid-infrared (mid-IR) band in the THz frequency range, spanning a much broader range in the electromagnetic spectrum, appear as promising candidates to unlock such limitations. Moreover, mid-IR bands are becoming options of interest due to their more stable performance through the atmospheric channel.

\begin{figure}[!htbp]
\centering
{\includegraphics[width=.8\textwidth]{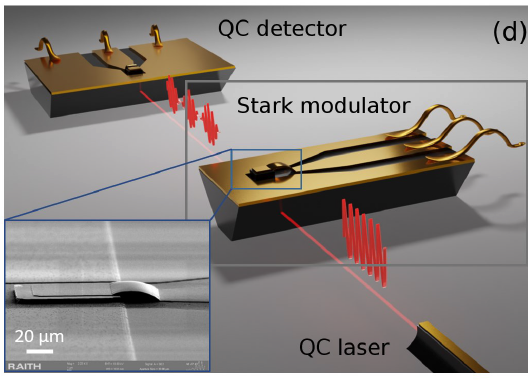}}
\caption{\em Sketch of the UQO system comprising QC laser, Stark modulator and QC detector, all of them operating at room temperature and at the same wavelength, 9 $\mu m$ (138 $meV$). The inset shows a scanning-electron-microscope image of the modulator connected via an air-bridge to the coplanar waveguide.}
\label{fig:UQOSystem}
\end{figure}

Optoelectronics technology with a very compact design ($\BigOSI{10}{\mu m}$), low power consumption ($\BigOSI{10}{mW}$), and operating at room temperature is well suited for HEP detectors. A recent work \cite{FreeSpaceDataTransmission} has demonstrated free optics transmission well above 1 Gb/s over a distance of 2 m. The unipolar quantum optoelectronics (UQO) system comprises three semiconductor devices operating in the mid-IR regime ($\lambda$=4-16 $\mu m$): a commercial continuous-wave quantum-cascade (QC) laser, a Stark modulator, and a QC detector (see Fig.~\ref{fig:UQOSystem}). The modulator is an asymmetric doped quantum well, where the energy shift between electrons in ground and excited states can be tuned in or out of resonance with the laser emission energy, under an applied bias. The modulation of the laser power absorbed can be directly controlled by the applied bias in the span of few volts. Therefore, a digital output of desired frequency can be created up to a cut-off frequency of 20 GHz.  The cut-off frequency is inversely proportional to the device surface (25$\times$25 $\mu m^2$). The charge modulator avoids implementation of gates for charge depletion, thus reducing intrinsic parasitic capacitance. In the QC detector absorbed photons excite electrons. After photo-excitation, electrons relax very rapidly by cascading towards the ground state and giving rise to a photocurrent. The electron relaxation time is estimated to be shorter than 10 ps: the intrinsic bandwidth is of the order of 100 GHz. However, the parasitic capacitance, induced by the mesa structure, limits the frequency range by more than one order of magnitude. The system has been fully characterized under different biases: good agreement with simulation is shown in terms of transmittance. As a possible application, the modulator was connected to a PRBS7 pulse generator and the corresponding eye of the modulated signal, which is received by the QC detector, is well open: a BER of 10$^{-12}$ is measured at 2.5 Gb/s. Given that 50-100 GHz bandwidth ranges for these devices are within reach, this technology is expected to approach Tb/s data rates, which together with the small footprints, makes it suitable for HEP detectors.

\section{Conclusion}
\label{s.conclusions} 

\par Novel link technologies are needed for readout of FCC-hh and muon collider experiments to cope with the higher data rates and radiation exposures. These experiments can also benefit from intelligent data reduction and aggregation techniques to optimize the readout bandwidth. Therefore, critical technologies include radiation-hard optical links, wireline, wireless, and free-space optics. It would be most efficient to develop them in close collaboration with commercial partners. Meanwhile, the HEP community will continue evaluating and adapting to emerging electronics and data processing technologies.



\bibliographystyle{JHEP.bst}
\bibliography{main}




\end{document}